\documentclass[letterpaper,12pt]{article}   
\usepackage{osajnl2} 
\usepackage[draft]{hyperref} 

\usepackage{amsmath}
\usepackage{amssymb}
\usepackage{stmaryrd}

\DeclareSymbolFont{AMSb}{U}{msb}{m}{n}
\DeclareMathSymbol{\N}{\mathbin}{AMSb}{"4E}
\DeclareMathSymbol{\Z}{\mathbin}{AMSb}{"5A}
\DeclareMathSymbol{\R}{\mathbin}{AMSb}{"52}
\DeclareMathSymbol{\Q}{\mathbin}{AMSb}{"51}
\DeclareMathSymbol{\I}{\mathbin}{AMSb}{"49}
\DeclareMathSymbol{\Cx}{\mathbin}{AMSb}{"43}

\newcommand{\At}{\ensuremath{\widetilde{A}}}
\newcommand{\Bt}{\ensuremath{\widetilde{B}}}
\newcommand{\gt}{\ensuremath{\widetilde{g}}} 
\newcommand{\Ct}{\ensuremath{\widetilde{C}}}
\newcommand{\Pc}{\ensuremath{\mathcal{P}}}

\begin{document}

\title{Analytical scattering matrix of subwavelength infinitely conducting metallic gratings in TM polarization}

\author{Philippe Boyer and Daniel Van Labeke}
\address{D\'{e}partement d'Optique P.M. Duffieux, Institut FEMTO-ST, CNRS UMR 6174
Universit\'{e} de Franche–Comt\'{e}, 25030 Besan\c{c}on Cedex, France}
\address{philippe.boyer@univ-fcomte.fr}

\begin{abstract}
We obtain in TM polarization an analytical expression of the scattering matrix of one infinitely conducting metallic lamellar grating with subwavelength slits. The theory is based on the Monomode Modal Method which consists in considering only one propagative mode in grating slits. Two expressions are exposed. The first one comes directly from the theory equations and the second one clearly reveals the Airy-like form of the scattering matrix terms. The theory is validated on a multi-grating object and the stability of the numerical results are shown at the same time. This work provides a basic and very efficient theoretical tool to calculate the diffraction by a stack of subwavelength metallic gratings.
\end{abstract}

\maketitle

\section{Introduction}
Since the discovery of the extraordinary optical transmission through subwavelength metallic gratings (SMG) \cite{art:Ebbesen,art:GarciaVidal10}, a lot of numerical methods \cite{art:Granet96,art:Popov00,art:Baida03,art:Bao05} have been applied to understand this singular physical phenomenon and to calculate the transmitted intensities. Now, it is known that such periodic sets of plasmonic or cavity resonators behave as Fabry-Perot interferometers. Today, the scientist's interest turns widely to applications, such as tunable optical filters \cite{art:Kim99,art:Pan05,art:Estruch11} or micro-polarizer exhibiting optical activity \cite{art:MacKay10,art:Plum09}. Some of these optical properties require complex structures made of a stack of SMG \cite{art:Shen04,art:Cheng08,art:Akiyama10,art:Xiang11} which often implies tricky and time-consuming computations. Thus, from fundamental physical considerations, it seems important to enhance the theoretical analysis in order to obtain very efficient computing codes to increase understanding of such structures and to easily discuss the influence of opto-geometrical parameters. 

For planar periodic objects like gratings surrounded by two semi-infinite homogeneous regions, the common and well-known theories such as the Classical Modal Method (CMM) \cite{art:Maystre72,art:Botten81}, the Rigorous Coupled-Wave Method (RCWM) \cite{art:Moharam81,art:Granet96,art:Lalanne96} or the Differential method (DM) \cite{art:Popov01,art:Neviere03}, usually lead to the calculation of two kinds of matrices relying the fields in homogeneous regions: the Scattering matrix (S-matrix) or the Transmission matrix (T-matrix). From them, iterative processes like the Scattering Matrix Propagation Algorithm (S-algorithm) \cite{art:Neviere03,art:Li96} are then used to compute the diffracted fields in multi-layer devices. For the case of gratings made of infinitely conducting metal, the CMM has remained for the last decades the most relevant method to solve the diffracting problem. It consists in considering the grating slits as waveguides and to express the cavity fields as a combination of waveguide modes. The initial formalism of the theory imposes to take as many modes (propagative or evanescent) in grating cavities as Rayleigh terms in homogeneous regions. But the transverse fields of evanescent modes make the method unstable (one matrix to invert becomes singular) when the slit width decreases and the number of harmonic terms increases. These numerical problems have limited its applications for a long time. By relevant matrix transformation, Gralak \emph{et al} have obtained non singular system to invert \cite{art:Gralak08}. 
The most studied configuration is when the cavity width is small enough to have only one propagative mode. The incident wavelength being larger than the cut-off of all the modes, except the fundamental one. When only one mode propagates in slits as in studied monoperiodic gratings lighted in TM polarization ($TM_0$ fundamental mode), this method has been named the Monomode Modal Method (MMM). This theory has been used before so far only to analytically express the propagative transmitted or reflected diffracted fields from which the diffracted efficiencies, transmittance and reflectance are deduced \cite{art:Lalane00,art:Boyer12,art:Rahman13}.
 
We propose to extend the MMM to the analytical calculation of the SMG S-matrix. It is worth noticing that the T-matrix cannot be analytically obtained since numerical inversion of one large matrix is required \cite{art:Shen04}. We only consider monoperiodic gratings lighted by a TM planewave. Therefore, the theory is restricted to thick gratings and wavelengths larger than the cut-off of the second cavity mode ($TM_1$). At first, we give the analytical expression of the SMG S-matrix directly deduced from the MMM equations. After that, we propose a more physical writing of the S-matrix terms from an Airy-like formulation which includes Fresnel coefficients at grating interfaces. Finally, the method is validated on a multi-grating device with the help of the S-algorithm and we verify that the numerical results are without numerical divergences.

\section{Scattering matrix of one subwavelength metallic grating}
\begin{figure}[t]
	\centerline{\includegraphics[scale=1.]{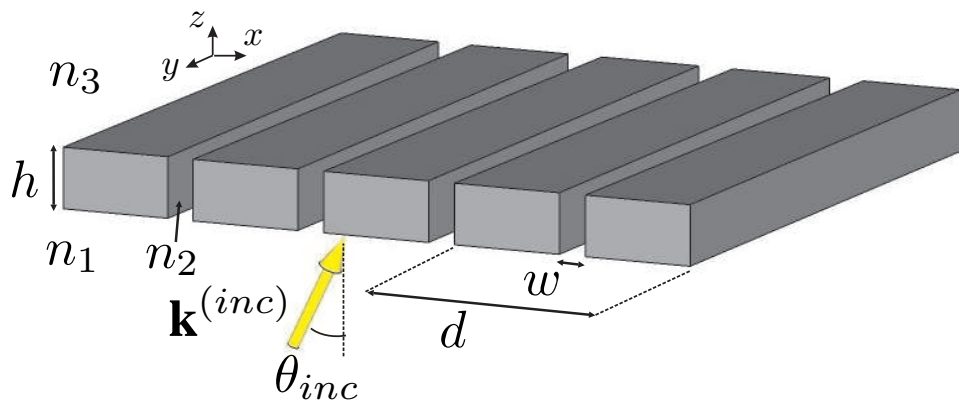}}
	\caption{\label{fig:multigratings}(Color online) Studied multilayer device: stack of metallic gratings.}
\end{figure}

We establish in this section the analytical expression of the S-matrix for one SMG surrounded by two semi-infinite homogeneous regions (see fig. \ref{fig:multigratings}). We assume that the metal is infinitly conducting and that refractive indices are noted $n_2$ for grating cavities and $n_j$ for homogeneous regions $j\in\left\{1,3\right\}$. The grating period is denoted $d$, its thickness $h=z_3-z_1$ and the cavity width $w$. In TM polarization, we know that the transverse fields are written as Fourier-Rayleigh (FR) expansions in homogeneous regions $(j)$: 
\begin{equation}
\textbf{E}_t^{(j)}=\textbf{e}_x\frac{e^{-i\omega t}}{\sqrt{d}}\sum_{p\in\Z}e^{i\alpha_p x}\left[A_p^{(j)}e^{i\gamma_p^{(j)}(z-z_j)}+B_p^{(j)}e^{-i\gamma_p^{(j)}(z-z_j)}\right]
\textrm{, }j\in\left\{1,3\right\}
\label{eq:ChHom}
\end{equation}
with $\gamma_p^{(j)}=\left[\epsilon_j\mu_0\omega^2-\alpha_p^2\right]^{1/2}$ and $\alpha_p=\alpha_0+p2\pi/d$. 

We note that the wavelength validity domain is $\left[\lambda_c,+\infty\right]$, where $\lambda_c=2w$ is the cut-off of the $TM_1$ mode and that the $TM_0$ mode has no cut-off. In fact, the cavity resonances related on high order modes ($TM_m$, $m\in\left[2,+\infty\right]$) cannot be excited since their resonance wavelengths are widely distant from this wavelength range. In the same way, their cut-off wavelengths are also widely smaller than the considered wavelengths since the grating is made of subwavelength cavities ($w<<\lambda$). Consequently, we can suppose that propagative and contra-propagative modes totally vanish during their evanescent propagation (specially for high $h$-values) and cannot establish stationnary resonances. For all these reasons, it seem judicious to consider in our theoretical analysis only the unique propagative mode $TM_0$. 

Consequently, the transverse field in cavities is written as the sum of propagative and contra-propagative $TM_0$ modes with amplitudes $\At$ and $\Bt$:
\begin{equation}
\textbf{E}_t^{(2)}=\textbf{e}_x\frac{e^{-i\omega t}}{\sqrt{w}}\left[\At e^{ik_0n_2(z-z_1)}+\Bt e^{-ik_0n_2(z-z_3)}\right]
\label{eq:ChCav}
\end{equation} 
with $k_0=2\pi/\lambda$. We emphasize that the choice of the phase origin in field expressions (\ref{eq:ChHom}) and (\ref{eq:ChCav}) is crucial to obtain an analytical expression of the S-matrix only depending on $h$. The projection of the field continuity relations for both electric and magnetic fields at grating interfaces on FR-orders and cavity mode transverse fields leads to $4$ equation sets \cite{art:Boyer12}:
\begin{equation}
A_p^{(3)}+B_p^{(3)}=\left(\At u+\Bt\right) \gt_p, \forall p\in\Z
\label{eq:projcontEsup}
\end{equation}
\begin{equation}
A_p^{(1)}+B_p^{(1)}=\left(\At+\Bt u\right) \gt_p, \forall p\in\Z
\label{eq:projcontEinf}
\end{equation}
\begin{equation}
\sum_{p=-\infty}^{+\infty}\eta_p^{(1)}\left[A_p^{(1)}-B_p^{(1)}\right]\gt_p^*=n_2\left(\At-\Bt u\right)
\label{eq:projcontHinf}
\end{equation}
\begin{equation}
\sum_{p=-\infty}^{+\infty}\eta_p^{(3)}\left[A_p^{(3)}-B_p^{(3)}\right]\gt_p^*=n_2\left(\At u-\Bt\right)
\label{eq:projcontHsup}
\end{equation}
where $u=exp(ik_0n_2h)$ with $k_0=2\pi/\lambda$, and the exponent $*$ denotes the complex conjugate. The overlap integrals $\gt_p$ between the cavity modes and FR-orders are: 
\begin{equation}
\gt_p=\sqrt{\frac{w}{d}}sinc\left(\alpha_p\frac{w}{2}\right)
\label{eq:defgp}
\end{equation}
The quantity $\eta_p^{(j)}=k_0\epsilon_j/\gamma_p^{(j)}$ is the admittance of FR $p$-order.

Injecting Eq. (\ref{eq:projcontEsup}) into (\ref{eq:projcontHsup}), then (\ref{eq:projcontEinf}) into (\ref{eq:projcontHinf}) leads to obtaining an equation set linking the mode amplitudes and the incident FR-orders. It writes in matricial form as:
\begin{equation}
\left(\begin{array}{cc}
\Ct^{(1)}+n_2 & \left[\Ct^{(1)}-n_2\right]u
\\\left[\Ct^{(3)}-n_2\right]u & \Ct^{(3)}+n_2
\end{array}\right)
\left(\begin{array}{c}
\At \\ \Bt
\end{array}\right)
=\left(\begin{array}{c}
2\sum_{p\in\Z}\eta_p^{(1)}\gt_p^*A_p^{(1)}
\\2\sum_{p\in\Z}\eta_p^{(3)}\gt_p^*B_p^{(3)}
\end{array}\right)
\label{eq:SystAB}
\end{equation}
where $\Ct^{(j)}$ are very important effective coefficients related to a sum of overlap integrals between the $TM_0$ mode and FR modes ponderated by the FR admittance:
\begin{equation}
\Ct^{(j)}=\sum_{p\in\Z}\eta_p^{(j)}\left|\gt_p\right|^2
\label{eq:defCj}
\end{equation}
As usual, the FR-expansions are in the following truncated into $2N+1$ harmonic terms in order to obtain matrices with finite size. Introducing column matrices $\left[A^{(j)}\right]$ and $\left[B^{(j)}\right]$ containing FR-amplitudes $A_p^{(j)}$ and $B_p^{(j)}$ respectively, the analytical inversion of the system (\ref{eq:SystAB}) which only requires inversion of a $2\times 2$ matrix, leads to simple relations relating $\At$ and $\Bt$ to all the $A_p^{(1)}$ and $B_p^{(3)}$ amplitudes:
\begin{equation}
\left(\begin{array}{c}
\At \\ \Bt
\end{array}\right)
=\Pc
\left(\begin{array}{c}
\left[A^{(1)}\right]
\\\left[B^{(3)}\right]
\end{array}\right)
\label{eq:matP}
\end{equation}
where $\Pc$ is a matrix of size $2\times (2N+1)$. The elements of its four blocks can be explicitely written: 
\begin{equation}
\left(\Pc_{11}\right)_p=\frac{2\left[n_2+\Ct^{(3)}\right]\eta_p^{(1)}\gt_p^*}{\left[\Ct^{(1)}+n_2\right]\left[\Ct^{(3)}+n_2\right]-u^2\left[\Ct^{(1)}-n_2\right]\left[\Ct^{(3)}-n_2\right]}
\label{eq:P11}
\end{equation}
\begin{equation}
\left(\Pc_{12}\right)_p=\frac{2u\left[n_2-\Ct^{(1)}\right]\eta_p^{(3)}\gt_p^*}{\left[\Ct^{(1)}+n_2\right]\left[\Ct^{(3)}+n_2\right]-u^2\left[\Ct^{(1)}-n_2\right]\left[\Ct^{(3)}-n_2\right]}
\label{eq:P12}
\end{equation}
\begin{equation}
\left(\Pc_{21}\right)_p=\frac{2u\left[n_2-\Ct^{(3)}\right]\eta_p^{(1)}\gt_p^*}{\left[\Ct^{(1)}+n_2\right]\left[\Ct^{(3)}+n_2\right]-u^2\left[\Ct^{(1)}-n_2\right]\left[\Ct^{(3)}-n_2\right]}
\label{eq:P21}
\end{equation}
\begin{equation}
\left(\Pc_{22}\right)_p=\frac{2\left[n_2+\Ct^{(1)}\right]\eta_p^{(3)}\gt_p^*}{\left[\Ct^{(1)}+n_2\right]\left[\Ct^{(3)}+n_2\right]-u^2\left[\Ct^{(1)}-n_2\right]\left[\Ct^{(3)}-n_2\right]}
\label{eq:P22}
\end{equation}
with $p\in\left[-N,N\right]$. Finally, an analytical expression of the S-matrix is basically obtained from Eqs. (\ref{eq:projcontEsup}) and (\ref{eq:projcontEinf}) by simply replacing $\At$ and $\Bt$ by their expressions given by Eq. (\ref{eq:matP}) and (\ref{eq:P11}) to (\ref{eq:P22}):
\begin{equation}
\left(\begin{array}{c}
\left[A^{(3)}\right] \\ \left[B^{(1)}\right]
\end{array}\right)
=S
\left(\begin{array}{c}
\left[A^{(1)}\right]
\\\left[B^{(3)}\right]
\end{array}\right)
\label{eq:matS}
\end{equation}
with 
\begin{equation}
\left(S_{11}\right)_{p,q}=\frac{4u\eta_p^{(1)}n_2\gt_p\gt_q^*}
{\left[\Ct^{(1)}+n_2\right]\left[\Ct^{(3)}+n_2\right]-u^2\left[\Ct^{(1)}-n_2\right]\left[\Ct^{(3)}-n_2\right]}
\label{eq:S11}
\end{equation}
\begin{equation}
\left(S_{12}\right)_{p,q}=2\eta_q^{(3)}\gt_p\gt_q^*\frac{\Ct^{(1)}+n_2+u^2\left[n_2-\Ct^{(1)}\right]}
{\left[\Ct^{(1)}+n_2\right]\left[\Ct^{(3)}+n_2\right]-u^2\left[\Ct^{(1)}-n_2\right]\left[\Ct^{(3)}-n_2\right]}
-\delta_{pq}
\label{eq:S12}
\end{equation}
\begin{equation}
\left(S_{21}\right)_{p,q}=2\eta_q^{(1)}\gt_p\gt_q^*\frac{\Ct^{(3)}+n_2+u^2\left[n_2-\Ct^{(3)}\right]}
{\left[\Ct^{(1)}+n_2\right]\left[\Ct^{(3)}+n_2\right]-u^2\left[\Ct^{(1)}-n_2\right]\left[\Ct^{(3)}-n_2\right]}
-\delta_{pq}
\label{eq:S21}
\end{equation}
\begin{equation}
\left(S_{22}\right)_{p,q}=\frac{4u\eta_p^{(3)}n_2\gt_p\gt_q^*}
{\left[\Ct^{(1)}+n_2\right]\left[\Ct^{(3)}+n_2\right]-u^2\left[\Ct^{(1)}-n_2\right]\left[\Ct^{(3)}-n_2\right]}
\label{eq:S22}
\end{equation}
with $(p,q)\in\left[-N,N\right]\times\left[-N,N\right]$ and $\delta_{pq}$ the Kr\"{o}necker symbol. The size of S-matrix is $2(2N+1)\times 2(2N+1)$. We notice that $\left(S_{22}\right)_{p,q}=\eta_q^{(1)}\left(S_{11}\right)_{p,q}/\eta_q^{(3)}$.

At this stage of our calculation, a rapid verification can be obtained. If only one incident FR-order is sent on the structure: $A_q^{(1)}=\delta_{q0}$ and $B_q^{(3)}=0$, in this case the diffracted amplitudes are the reflection and transmission coefficients: $A_p^{(3)}=t_p$ and $B_p^{(1)}=r_p$. Their analytical expression are given by the relation $t_p=\left(S_{11}\right)_{p,0}$ and $r_p=\left(S_{21}\right)_{p,0}$ which coincide with published ones \cite{art:Boyer12}. In the same way, $\At=\left(\Pc_{11}\right)_0$ and $\Bt=\left(\Pc_{21}\right)_0$ provide the expresion of the cavity mode amplitudes.

\section{Another expression of S-matrix}
\begin{figure}[t]
	\centerline{\includegraphics[scale=0.6]{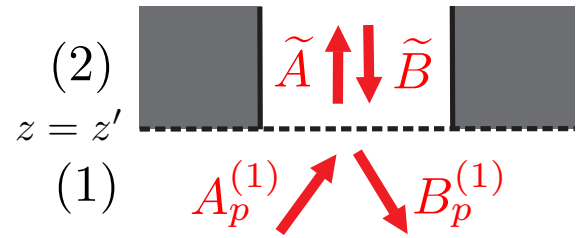}}
	\caption{\label{fig:fresnel}(Color online) Schematic represantation of interface between an homogeneous region and a metallic waveguide for Fresnel coefficient calculation. Only the $TM_0$ mode is considered in the waveguide.}
\end{figure}

We can obtain a more physical expression of the S-matrix which lets clearly appear the reflection and transmission coefficients at interfaces. At first, we have to explicit the Fresnel coefficients at the interfaces between a homogeneous region $(1)$ and a metallic waveguide $(2)$ as shown in fig. \ref{fig:fresnel}. The fields are expressed as in eq. (\ref{eq:ChHom}) and (\ref{eq:ChCav}) but $z_1=z_3=z'$, the interface coordinate on $z$-axis. We introduce the S-matrix of the interface:
\begin{equation}
\left(\begin{array}{c}
A^{(2)} \\ \left[B^{(1)}\right]
\end{array}\right)
=\left(\begin{array}{cc}
t^{(1,2)} & r^{(2,1)}
\\r^{(1,2)} & t^{(2,1)}
\end{array}\right)
\left(\begin{array}{c}
\left[A^{(1)}\right]
\\B^{(2)}
\end{array}\right)
\label{eq:matSint}
\end{equation}
Noticing $N$ the truncation order of FR-expansions, we highlight that $t^{(1,2)}$ is a line $1\times (2N+1)$-block, $r^{(1,2)}$ is a square $(2N+1)\times (2N+1)$-block, $r^{(2,1)}$ is a scalar block and $t^{(2,1)}$ is a column $(2N+1)\times 1$-block. We can easily prove that their elements are expressed by:
\begin{equation}
t_p^{(1,2)}=\frac{2\eta_p^{(1)}\gt_p^*}{n_2+\Ct^{(1)}}
\end{equation}
\begin{equation}
r_{p,q}^{(1,2)}=t_q^{(1,2)}\gt_p-\delta_{pq}
\end{equation}
\begin{equation}
r^{(2,1)}=\frac{n_2-\Ct^{(1)}}{n_2+\Ct^{(1)}}
\end{equation}
\begin{equation}
t_p^{(2,1)}=\left[1+r_p^{(2,1)}\right]\gt_p=\frac{2n_2\gt_p}{n_2+\Ct^{(1)}}
\end{equation}
with $p\in\left[-N,N\right]$ and $q\in\left[-N,N\right]$. We see that $t_p^{(1,2)}$ and $r^{(2,1)}$ are defined as usual amplitude ratios: $t_p^{(1,2)}=\frac{B_p^{(1)}}{\Bt}$ and $r^{(2,1)}=\frac{\At}{\Bt}$ without incident $q$-order. But $t_p^{(2,1)}$ and $r_{p,q}^{(1,2)}$ are partial reflection and transmission coefficients. For instance, $r_{p,q}^{(1,2)}$ expresses the reflected $q$-order when only the incident $p$-order falls on the gratings. A summation is necessary otherwise (several incident $p$-orders).

The S-matrix of one metallic grating is obtained by applying twice the S-algorithm between the two S-matrices of each interfaces and the S-matrix related to the cavity mode propagation. We repeat that it is an iterative algorithm which consists in calculating de S-matrix $S^{(c)}$ of two adjacent layers according to their S-matrices $S^{(a)}$ and $S^{(b)}$ \cite{art:Li96,art:Montiel97}:
\begin{equation}
S_{11}^{(c)}=S_{11}^{(b)}\left[I_d+S_{12}^{(a)}ZS_{21}^{(b)}\right]S_{11}^{(a)}
\label{eq:SalgS11}
\end{equation}
\begin{equation}
S_{12}^{(c)}=S_{11}^{(b)}S_{12}^{(a)}ZS_{22}^{(b)}+S_{12}^{(b)}
\label{eq:SalgS12}
\end{equation}
\begin{equation}
S_{21}^{(c)}=S_{22}^{(a)}ZS_{21}^{(b)}S_{11}^{(s)}+S_{21}^{(a)}
\label{eq:SalgS21}
\end{equation}
\begin{equation}
S_{22}^{(c)}=S_{22}^{(a)}ZS_{22}^{(b)}
\label{eq:SalgS22}
\end{equation}
with
\begin{equation}
Z=\left[I_d-S_{21}^{(b)}S_{12}^{(a)}\right]^{-1}
\label{eq:SalgZ}
\end{equation}
where $I_d$ denotes the identity matrix. Before applying the S-algorithm, we remark that the phase origin of the cavity fields chosen in eq. (\ref{eq:ChCav}) is not the same as the one used to explicit the S-matrix of one interface in eq. (\ref{eq:matSint}). Thus, we have to replace $\Bt$ by $\Bt u$ in eq. (\ref{eq:matSint}) for the matrix $S^{(1)}$ at $z=z_1$. In the same way, $\At$ is changed to $\At u$ for the matrix $S^{(3)}$ at $z=z_3$. The mode propagation S-matrix $S^{(2)}$ in cavities is obviously equal to $uI_d$. The S-algorithm is first applied on $S^{(1)}$ and $S^{(2)}$ which leads to the matrix $S^{(1,2)}$:
\begin{equation}
S^{(1,2)}=\left(\begin{array}{cc}
ut^{(1,2)} & u^2r^{(2,1)}
\\r^{(1,2)} & ut^{(2,1)}
\end{array}\right)
\end{equation}
A second application of S-algorithm on $S^{(1,2)}$ and $S^{(3)}$ leads to the S-matrix $S^{(1,3)}=S$ of the grating. Its blocks have a simple expression:
\begin{equation}
\left(S_{11}\right)_{p,q}=\frac{t_p^{(2,3)}t_q^{(1,2)}u}{1-r^{(2,1)}r^{(2,3)}u^2}
\label{eq:S11f}
\end{equation}
\begin{equation}
\left(S_{12}\right)_{p,q}=\frac{\gt_pt_q^{(3,2)}\left[1+r^{(2,1)}u^2\right]}{1-r^{(2,1)}r^{(2,3)}u^2}-\delta_{pq}
\label{eq:S12f}
\end{equation}
\begin{equation}
\left(S_{21}\right)_{p,q}=\frac{\gt_pt_q^{(1,2)}\left[1+r^{(2,3)}u^2\right]}{1-r^{(2,1)}r^{(2,3)}u^2}-\delta_{pq}
\label{eq:S21f}
\end{equation}
\begin{equation}
\left(S_{22}\right)_{p,q}=\frac{t_p^{(2,1)}t_q^{(3,2)}u}{1-r^{(2,1)}r^{(2,3)}u^2}
\label{eq:S22f}
\end{equation}
This formulation is equivalent to the one given in eqs. (\ref{eq:S11}) to (\ref{eq:S22}) but clearly reveals that the S-matrix terms can be expressed as Airy-like formulae. We also prove that the $\Pc$-matrix can be written:
\begin{equation}
\left(\Pc_{11}\right)_p=\frac{t_p^{(1,2)}}{1-r^{(2,1)}r^{(2,3)}u^2}
\label{eq:P11f}
\end{equation}
\begin{equation}
\left(\Pc_{12}\right)_p=\frac{r^{(2,1)}t_p^{(3,2)}u}{1-r^{(2,1)}r^{(2,3)}u^2}
\label{eq:P12f}
\end{equation}
\begin{equation}
\left(\Pc_{21}\right)_p=\frac{r^{(2,3)}t_p^{(1,2)}u}{1-r^{(2,1)}r^{(2,3)}u^2}
\label{eq:P21f}
\end{equation}
\begin{equation}
\left(\Pc_{22}\right)_p=\frac{t_p^{(3,2)}}{1-r^{(2,1)}r^{(2,3)}u^2}
\label{eq:P22f}
\end{equation}

\section{A convergence test}
\begin{figure}[t]
	\centerline{\includegraphics[scale=1.]{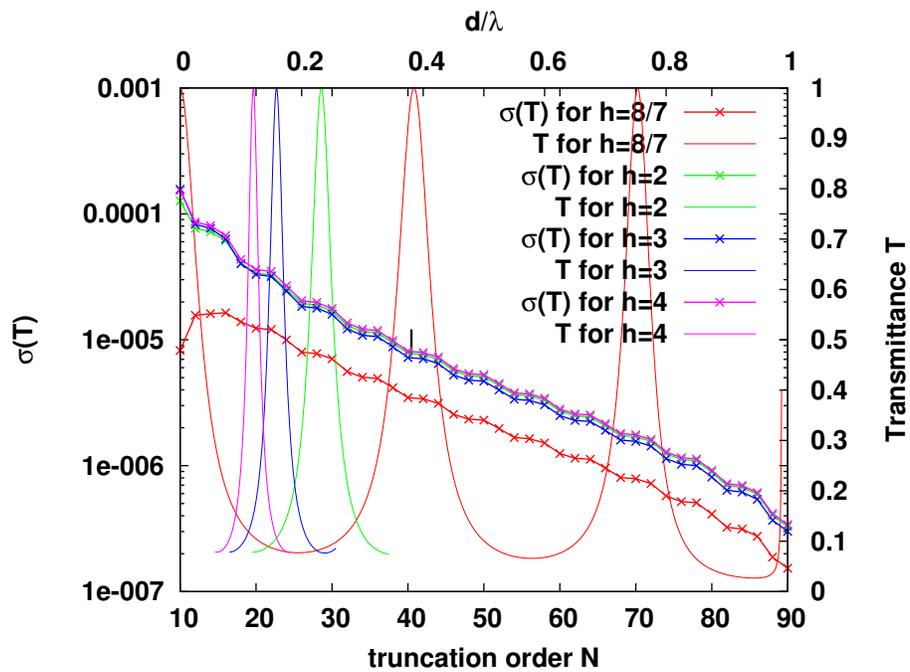}}
	\caption{\label{fig:convN}(Color online) Convergence test of $\sigma(T)$ according to the truncation Fourier-Rayleigh order $N$ and for different values of $h$. $\sigma(T)$ is evaluated at the first order resonance peak in shown transmittance curves versus $\lambda$.}
\end{figure}

Before validating the theory on multilayer devices, we have analyzed the numerical stability of the analytical S-matrix computation on a single SMG. Many numerical methods based on similar formalisms to the CMM, the RCWM or the DM are fundamentally unstable. Precisely, they often diverge when the truncation order of FR-expansions $N$ and/or the grating thickness increase. The S-algorithm and the Fast Fourier Factorization \cite{art:Popov01,art:Neviere03} bring solutions to these numerical problems. In order to evaluate the stability of our method, we have to define a convergence criteria. For every value of $N$ and $h$, the computed quantity $R+T$ is exactly equal to $1$ (energy balance criteria) and so cannot be chosen to estimate convergence. Thus, we introduce the relative accuracy $\sigma(N)=[T(N)-T(N_{max})]/T(N_{max})$, between the transmittance value at $N$ and the one at $N_{max}=100$ ($N\leq N_{max}$). The numerical analysis is done on the same structures studied in \cite{art:Shen04}. The figure \ref{fig:convN} shows $\sigma(N)$ according to $N$ for different values of $h$ and when the transmittance is evaluated to the first peak maxima (at $d/\lambda=0.385$ for $h=8/7$). The transmittance versus $d/\lambda$ is also plotted in fig. \ref{fig:convN} for every $h$ value (plotted only on the wavelength range of the first peak for $h=2$ to $4$). The other parameters are fixed to $d=1$, $w=1/7$, $n_1=n_2=n_3=1$ and $\theta_{inc}=0^o$.  We see that the numerical results converge and remain stable whatever the $h$ value. In fact, it is shown that the numerical instabilities in the classical modal method result from vanishing overlap integrals between high-order cavity modes and high-order FR-amplitudes, which leads to non-invertible matrices. Gralak \emph{et al} \cite{art:Gralak08} have proposed a solution to avoid such numerical problems by considerations on the matrix form. Stable formalism can be intrinsically obtained by considering only the propagating cavity mode as in our equations. The convergence to zero of the overlap integral given by eq. (\ref{eq:defgp}) does not affect the stability of the analytical $S$-matrix terms: The coupling coefficients $\Ct^{(j)}$ defined in eq. (\ref{eq:defCj}) and expressed from overlap integrals appear in the denominator of eqs. (\ref{eq:S11}) to (\ref{eq:S22}) and never tends to zero (but its modulus can tends to infinity at the Rayleigh wavelengths \cite{art:Boyer12}). More generally, it is proven that the denominator in $S$-matrix terms vanishes only at resonance conditions when the free oscillation problem is solved (searching of complex resonance frequencies) \cite{art:Boyer12}.

\section{A multilayered structure}
The numerical analysis of the diffraction by a multilayer device necessarily requires the calculation of its S-matrix. The use of the iterative S-algorithm (see eq.(\ref{eq:SalgS11}) to (\ref{eq:SalgZ})) allows the computation of this global S-matrix of the studied structure from the S-matrices of each layer. The number of layers is denoted $L$. One layer can be a metallic grating (see fig. \ref{fig:multigratings}) or a homogeneous region as for the example of stacks of metallic gratings separated by homogeneous regions. Thus, it is necessary to explicit the S-matrix of a homogeneous layer. The fields in the homogeneous cavity are written as FR-expansions. The $\Pc$-matrix of a homogeneous layer is a $4\times 4$ diagonal block matrix. Its elements are expressed as in eqs. (\ref{eq:P11f}) to (\ref{eq:P22f}) but with $u\equiv u_p$ and $r^{(2,j)}\equiv r_p^{(2,j)}$, $j\in\left\{1,3\right\}$. All the reflection and transmission terms are the classical Fresnel coefficients for FR-orders:
\begin{equation}
r_p^{(1,2)}=\frac{\eta_p^{(1)}-\eta_p^{(2)}}{\eta_p^{(1)}+\eta_p^{(2)}}
\label{eq:r12hom}
\end{equation}
\begin{equation}
t_p^{(1,2)}=\frac{2\eta_p^{(1)}}{\eta_p^{(1)}+\eta_p^{(2)}}
\label{eq:t12hom}
\end{equation}
where $(1)$ and $(2)$ refer to the two homogeneous regions on both sides of one interface. The $S$-matrix is also a $4\times 4$ diagonal block matrix. Its elements are expressed as in eqs. (\ref{eq:S11f}) to (\ref{eq:S22f}) but with $\gt_p\equiv 1$, $(P_{k,l})_{p,q}=(P_{k,l})_p\delta_{p,q}$, $(k,l)\in\left\{1,2\right\}$ and with notations introduced in eqs. (\ref{eq:r12hom}) and (\ref{eq:t12hom}).

\begin{figure}[t]
	\centerline{\includegraphics[scale=1.]{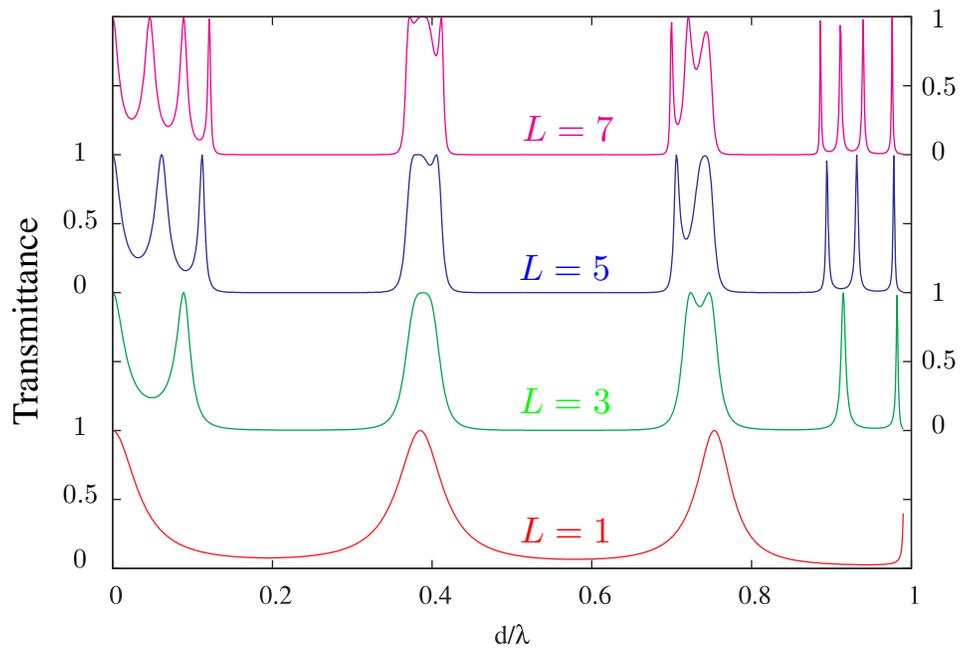}}
	\caption{\label{fig:multigratingsT}(Color online) Transmittance versus wavelength for multilayer devices. $L$ denotes the number of layers. $L=1$ for one grating ; $L=3$ for two gratings and one homogeneous layer, etc.}
\end{figure}

We validate our theory on a stack of identical metallic gratings lighted in normal incidence. We consider the same grating as the previous one. The gratings are separated by identical homogeneous layers of thickness $h_{hom}=4/7$ and filled with air. We have studied four structures from $L=1$ to $L=7$ where $L=2L_g-1$ is the total number of layers and $L_g\in\N$ is the number of metallic gratings (there are $L_g-1$ homogeneous layers). $N$ is fixed to $20$. The transmittances are plotted in fig. \ref{fig:multigratingsT}. We obviously find again the transmittance curve for one and four gratings as studied in \cite{art:Shen04}. We also verify that new resonance peaks appear when a grating is added. In fact, some peaks are due to degeneracy splitting by coupling of two surrounded gratings, each of them behaving as a resonator. Other peaks are explained by the resonances of the homogeneous cavity created between two gratings and behaving as a Fabry-Perot resonator.

\section{Conclusion}
We have obtained in TM polarization an analytical form for the S-matrix of a lamellar metallic grating because only one cavity mode is taken into account. In fact, the $2\times 2$-matrix which links propagative and contra-propagative $TM_0$ cavity modes can easily be inverted. As example, considering two propagative modes induces a tricky analytical inversion of a $4\times 4$-matrix. A generalization to a finite number of propagative slit modes leads to a semi-analytical formulation that we will present in a future work. We have also verified that the theory is numerically stable: the computations do not diverge when the number of Fourier-Rayleigh harmonics and the grating thickness increase. The main interest of the method remains the study of multilayer systems. It has been validated on a stack of metallic gratings. 

The theory may also be extended to the $TE$ case but its wavelength validity domain is restricted to the wavelength interval for which one propagative mode exists, i.e. between $TE_0$ and $TE_1$ mode cut-off wavelengths. For the $TM$ case, we restipulate that the $TM_0$ mode has no cut-off which lends huge validity domain.

In our future work, we shall apply this formalism to bi-periodic devices in order to develop nanostructures with polarization effects like, for example, an optical activity. We shall in addition present the semi-analytical calculation of the stable scattering matrix valid for metallic gratings with arbitrary cavity width including several propagative modes.

\vspace{1cm}
\emph{Acknowledgements:} I would like to thank Donna L'H\^{o}te from the Centre de linguistique appliqu\'{e}e (CLA) of Besan\c{c}on (France) for her helpful advice.


\clearpage
\section*{List of Figure Captions}
\noindent Fig. 1. (Color online) Studied multilayer device: stack of metallic gratings.

\vspace{0.2cm}
\noindent Fig. 2. (Color online) Schematic represantation of interface between an homogeneous region and a metallic waveguide for Fresnel coefficient calculation. Only the $TM_0$ mode is considered in the waveguide.

\vspace{0.2cm}
\noindent Fig. 3. (Color online) Convergence test of $\sigma(T)$ according to the truncation Fourier-Rayleigh order $N$ and for different values of $h$. $\sigma(T)$ is evaluated at the first order resonance peak in shown transmittance curves versus $\lambda$.

\vspace{0.2cm}
\noindent Fig. 4. (Color online) Transmittance versus wavelength for multilayer devices. $L$ denotes the number of layers. $L=1$ for one grating ; $L=3$ for two gratings and one homogeneous layer, etc.

\clearpage

\clearpage

\clearpage

\clearpage

\end{document}